\documentclass[english,11pt]{article}
\usepackage[T1]{fontenc}
\usepackage{booktabs}
\usepackage{amsmath}
\usepackage{graphicx}
\usepackage{wasysym}

\makeatletter

\providecommand{\tabularnewline}{\\}

\newcommand{\lyxaddress}[1]{
\par {\raggedright #1
\vspace{1.4em}
\noindent\par}
}

\usepackage{mathrsfs}   %\mathscr{L}
\usepackage{slashed}     %\slashed{p}
\usepackage{url} \usepackage{graphicx}
\usepackage[colorlinks=true,linkcolor=redLinks,citecolor=greenLinks,urlcolor=redLinks,
pdfborder={0 0 1}]{hyperref} \usepackage{xcolor} \usepackage{framed}
\usepackage[numbers,sort&compress]{natbib} \usepackage{amsmath}
 \def\lfv{lepton flavour
  violation } \def\lnv{lepton number violation }
\def\vev#1{\left\langle #1\right\rangle}   \allowdisplaybreaks

\colorlet{shadecolor}{gray!15}

\definecolor{greenLinks}{rgb}{0, 0.6, 0} 
\definecolor{blueLinks}{rgb}{0, 0, 0.6}
\definecolor{redLinks}{rgb}{0.6, 0, 0}
\definecolor{tempText}{rgb}{0.55, 0.10,0.67}
\definecolor{eprintLinks}{rgb}{0.4, 0.4, 0.4}
\definecolor{journalLinks}{rgb}{0.6, 0, 0}

\newcommand{\MYhref}[3][redLinks]{\href{#2}{\color{#1}{#3}}}%

\usepackage{multirow}
\let\orig@Hy@EveryPageAnchor\Hy@EveryPageAnchor
\def\Hy@EveryPageAnchor{%
    \begingroup
    \hypersetup{pdfview=Fit}%
    \orig@Hy@EveryPageAnchor
    \endgroup
}

\let\oldFootnote\footnote
\newcommand\nextToken\relax

\newcommand{\sm}{{standard model }}
\renewcommand\footnote[1]{%
    \oldFootnote{#1}\futurelet\nextToken\isFootnote}

\newcommand\isFootnote{%
    \ifx\footnote\nextToken\textsuperscript{,}\fi}

\makeatother

\usepackage{babel}
\begin{document}

% \title{{\Large{}\vspace{-1.0cm}} \hfill {\normalsize{}IFIC/15-XX}
%   \\*[10mm] Stability of triplet seesaw model
%   revisited{\Large{}\vspace{0.5cm}}}

\title{{\Large{}\vspace{-1.0cm}} \hfill %%{\normalsize{}IFIC/15-XX}
  \\*[10mm] Consistency of the triplet seesaw model
  revisited{\Large{}\vspace{0.5cm}}}

\author{{\Large{}Cesar Bonilla}\thanks{E-mail: cbonilla@ific.uv.es} {\Large{},
Renato M. Fonseca}\thanks{E-mail: renato.fonseca@ific.uv.es} , {\Large{}J.
W. F. Valle}\thanks{E-mail: valle@ific.uv.es} \date{}}

\maketitle

\lyxaddress{\begin{center}
{\Large{}\vspace{-0.5cm}}AHEP Group, Instituto de F\'isica Corpuscular,
C.S.I.C./Universitat de Val\`encia\\
Edificio Institutos de Investigaci\'on, Apartado 22085, E--46071 Valencia,
Spain
\par\end{center}}

\begin{center}
\today
\par\end{center}
\begin{abstract}
  Adding a scalar triplet to the Standard Model is one of the simplest
  ways of giving mass to neutrinos, providing at the same time a
  mechanism to stabilize the theory's vacuum. In this paper, we
  revisit these aspects of the type-II seesaw model pointing out that
  the bounded-from-below conditions for the scalar potential in use in
  the literature are not correct. We discuss some scenarios where the
  correction can be significant and sketch the typical scalar boson
  profile expected by consistency.
\end{abstract}

\section{Introduction}

More than ever, after the discovery of the Higgs boson, particle
physicists are eager for new results that can shed light on the
symmetry breaking puzzle.
The tiny neutrino masses suggest that probably a different mass
generation scheme associated to their charge neutrality is at work.
Neutrino masses can be introduced in the Standard Model (SM) through
the lepton number violating coupling of a scalar triplet $\Delta$ 
(hypercharge $+1$) with the left-handed leptons,
\begin{equation}
\frac{Y_{\Delta,ij}}{2}L_{i}^{T}C\left(i\tau_{2}\right)\Delta L_{j}+\textrm{h.c.}\label{eq:1}
\end{equation}
and generate a neutrino mass matrix $Y_{\Delta}\vev{\Delta^{0}} $
after electroweak symmetry breaking. Here $i\tau_2$ is the weak isospin
conjugation matrix. The vacuum expectation value of the triplet is
proportional to the strength $m_{H\Delta}$ of the coupling $HH\Delta$
which can be an arbitrarily small parameter since this is the only
lepton number violating coupling in the model. This is arguably the
most economical way of realizing Weinberg' s dimension five
operator~\cite{weinberg:1980bf}. For simplicity here we focus upon the
case of explicit \lnv~\cite{Schechter:1980gr} since the implementation
of spontaneous \lnv~\cite{Schechter:1981cv} would require an extended
scalar sector containing also a singlet. In this scheme one
``explains'' the smallness of neutrino masses with the smallness of
$m_{H\Delta}$ --- and hence the smallness of the ``induced'' vacuum
expectation value (VEV) $v_\Delta \equiv \vev{\Delta^{0}}$ --- even
with a light messenger scalar triplet $\Delta$, potentially accessible
at the next run of the LHC.

On the other hand, it is known that the Higgs quartic coupling in the
SM is driven to negative values at high energies, before the Planck
scale is reached~\cite{Sher:1988mj,Alekhin:2012py}. With the triplet
scalar field, the situation changes as the new quartic scalar
interactions between $H$ and $\Delta$ are able to soften the decrease
of the Higgs quartic coupling $\lambda_{H}$ as the energy scale is
increased \cite{Gogoladze2008,Chao2012,Chun:2012jw,Dev:2013ff}.  The
effect is qualitatively the same if the triplet is replaced by an
$SU(2)_{L}$ singlet
\cite{Lebedev2012,Elias-Miro2012,Falkowski2015,Costa:2014qga,Bonilla2015}.
However, with the new triplet scalar, it is no longer enough to check
that the Higgs quartic coupling stays positive, as the conditions for
the potential to be bounded from below become more elaborate.

Regardless of the energy scale one may ask, under what conditions is
the potential of the type-II seesaw model bounded from below? An attempt
to write down for the first time these necessary and sufficient vacuum 
stability conditions taking into account all field directions has been 
made in \cite{Arhrib:2011uy}. However, as we point
out in this paper, those conditions are too strong --- they are
sufficient but not necessary to ensure that a set of values for the
quartic couplings corresponds to a stable vacuum.
The structure of this paper is the following: after a brief review of
the basic properties of the model (section \ref{sec:2}) we derive the
necessary and sufficient conditions for the potential to be bounded
from below in section \ref{sec:3}, discussing the difference with the
conditions in use in the literature both from a theoretical
point-of-view as well as a numerical one. In section
\ref{sec:Regions-of-stability} we apply these conditions to explore
the region in parameter space of the type-II seesaw where the
potential is stable up to some given scale.  Finally, we present some
conclusions in section \ref{sec:Final-remarks}.  (Two appendices
provide supplementary material.)

\section{\label{sec:2}Basic properties of the type-II seesaw model}

Here we consider the simplest neutrino mass generation scheme based on
an effective seesaw mechanism with explicit \lnv described by the complex
triplet, given as
\begin{align}
\Delta\equiv\left(\begin{array}{cc}
\frac{\Delta^{+}}{\sqrt{2}} & \Delta^{++}\\
\Delta^{0} & -\frac{\Delta^{+}}{\sqrt{2}}
\end{array}\right) & \,.
\end{align}
The most general potential
involving $\Delta$ and the Standard Model Higgs doublet
$H=\left(H^{+},H^{0}\right)^{T}$ has a total of eight parameters which
we can take to be real:
\begin{align}
V\left(H,\Delta\right) & =-\mu_{H}^{2}H^{\dagger}H+\mu_{\Delta}^{2}\textrm{Tr}\left(\Delta^{\dagger}\Delta\right)+\left[\frac{m_{H\Delta}}{2}H^{T}\left(i\tau_{2}\right)\Delta^{\dagger}H+\textrm{h.c.}\right]+\frac{1}{2}\lambda_{H}\left(H^{\dagger}H\right)^{2}\nonumber \\
 & +\lambda_{H\Delta}\textrm{Tr}\left(\Delta^{\dagger}\Delta\right)\left(H^{\dagger}H\right)+\lambda'_{H\Delta}H^{\dagger}\Delta\Delta^{\dagger}H+\frac{\lambda_{\Delta}}{2}\left[\textrm{Tr}\left(\Delta^{\dagger}\Delta\right)\right]^{2}+\frac{\lambda'_{\Delta}}{2}\textrm{Tr}\left(\Delta^{\dagger}\Delta\Delta^{\dagger}\Delta\right)\,.\label{eq:potential}
\end{align}
The vacuum expectation value of the neutral component of the triplet,
$v_{\Delta}\equiv \vev{ \Delta^{0}} $, must be significantly smaller
than the one of the standard Higgs, $v_{H}\equiv\vev{ H^{0}} $,
otherwise the $\rho$ parameter will deviate too much from 1. Indeed,
\begin{align}
\rho & \approx1-2\alpha^{2}
\end{align}
with $\alpha\equiv v_{\Delta}/v_{H}$ so this ratio of VEVs can be at
most of the percent order given the experimental constraints on
$\rho$~\cite{Agashe:2014kda}. Furthermore, since neutrino masses are
proportional to $ v_{\Delta}$, this VEV should indeed be very small.
Under the approximation that $\alpha\ll1$, the minimization solution
of the potential requires that
\begin{align}
\mu_{H}^{2} & \approx\lambda_{H}v_{H}^{2}\,,\\
\mu_{\Delta}^{2} & \approx\left(\frac{\chi}{2}-\lambda_{H\Delta}-\lambda'_{H\Delta}\right)v_{H}^{2}\,,
\end{align}
where 
\begin{equation}
  \label{eq:chi}
\chi\equiv m_{H\Delta}/v_{\Delta} . 
\end{equation}
Using these relations one can write the scalar boson mass eigenstates
as shown in table~\ref{tab:Mass-eigenstates}.

\begin{table}[tbph]
\begin{centering}
\begin{tabular}{ccc}
\toprule 
Mass eigenstate $\phi$ & Mass squared $m_{\phi}^{2}$ & Composition\tabularnewline
\midrule
$H^{++}$ & $v_{H}^{2}\left(\frac{\chi}{2}-\lambda'_{H\Delta}\right)$ & $\Delta^{++}$\tabularnewline
$G^{+}$ & $0$ & $H^{+}+\sqrt{2}\alpha\Delta^{+}$\tabularnewline
$H^{+}$ & $v_{H}^{2}\left(\frac{\chi}{2}-\frac{\lambda'_{H\Delta}}{2}\right)$ & $\Delta^{+}-\sqrt{2}\alpha H^{+}$\tabularnewline
$G^{0}$ & $0$ & $H_{I}^{0}+2\alpha\Delta_{I}^{0}$\tabularnewline
$A^{0}$ & $\frac{1}{2}v_{H}^{2}\chi$ & $\Delta_{I}^{0}-2\alpha H_{I}^{0}$\tabularnewline
$h^{0}$ & $2\lambda_{H}v_{H}^{2}$ & $H_{R}^{0}+2\alpha\frac{\chi-2\lambda_{H\Delta}-2\lambda'_{H\Delta}}{\chi-4\lambda_{H}}\Delta_{R}^{0}$\tabularnewline
$H^{0}$ & $\frac{1}{2}v_{H}^{2}\chi$ & $\Delta_{R}^{0}-2\alpha\frac{\chi-2\lambda_{H\Delta}-2\lambda'_{H\Delta}}{\chi-4\lambda_{H}}H_{R}^{0}$\tabularnewline
\bottomrule
\end{tabular}
\par\end{centering}
\protect\caption{\label{tab:Mass-eigenstates}Scalar mass eigenstates in the type-II seesaw
  model. We have defined the dimensionless parameters
 $\alpha\equiv v_{\Delta}/v_{H}$ and 
 $\chi\equiv m_{H\Delta}/v_{\Delta}$.}
\end{table}
Note that if the doubly charged Higgs $H^{++}$ is to be heavier than
half a TeV or so, then $\chi\apprge10$, making $\chi$ significantly
larger than any of the quartic couplings $\lambda_{i}$ which one
expects to be, at most, of order 1. Moreover, one sees that for a
suitable $\chi$ the would-be triplet Nambu-Goldstone boson state
$A^{0}$ can be massive enough to have escaped detection at LEP.

\section{\label{sec:3}When is the scalar potential bounded from below?}

We now turn to the important issue of the stability of the VEV
solution mentioned above. As long as all scalar masses are positive,
the potential will not roll down classically to another minimum, but
this still leaves open the possibility of a tunneling to a deeper
minimum. In order for this not to happen, it is necessary (although
not sufficient) that the potential does not fall to infinitely
negative values in any VEV direction. In other words, we must ensure
that $V$ is bounded from below, which is equivalent to the requirement
that the quartic part of the potential in equation
\eqref{eq:potential}, $V^{(4)}$, must be positive for all non-zero
field values.  In the following then, we shall derive the necessary
and sufficient conditions for this to be true, correcting the result
obtained in~\cite{Arhrib:2011uy}.

While there are ten real degrees of freedom (four in $H$ plus six in
$\Delta$), $V$ depends on them only through 4 quantities:
$H^{\dagger}H$, $\textrm{Tr}\left(\Delta^{\dagger}\Delta\right)$,
$H^{\dagger}\Delta\Delta^{\dagger}H$ and
$\textrm{Tr}\left(\Delta^{\dagger}\Delta\Delta^{\dagger}\Delta\right)$.
In the following, we shall take
$\textrm{Tr}\left(\Delta^{\dagger}\Delta\right)$ to be non-zero.%
\footnote{If this is not the case, the quartic part of the potential is reduced to
$\frac{1}{2}\lambda_{H}\left(H^{\dagger}H\right)^{2}$ in which case it
is clear that one must have $\lambda_{H}>0$.} We now define $r$, $\zeta$
and $\xi$ as the following non-negative dimensionless quantities
\cite{Arhrib:2011uy},
\begin{align}
H^{\dagger}H & \equiv r\textrm{Tr}\left(\Delta^{\dagger}\Delta\right)\,,\label{eq:definition_r}\\
\textrm{Tr}\left(\Delta^{\dagger}\Delta\Delta^{\dagger}\Delta\right) & \equiv\zeta\left[\textrm{Tr}\left(\Delta^{\dagger}\Delta\right)\right]^{2}\,,\label{eq:definition_rho}\\
H^{\dagger}\Delta\Delta^{\dagger}H & \equiv\xi\textrm{Tr}\left(\Delta^{\dagger}\Delta\right)\left(H^{\dagger}H\right)\,,\label{eq:definition_phi}
\end{align}
such that the quartic part of the potential reads
\begin{align}
\frac{V^{(4)}}{\left[\textrm{Tr}\left(\Delta^{\dagger}\Delta\right)\right]^{2}} & =\frac{1}{2}\lambda_{H}r^{2}+\lambda_{H\Delta}r+\lambda'_{H\Delta}\xi r+\frac{\lambda_{\Delta}}{2}+\frac{\lambda'_{\Delta}}{2}\zeta\,.\label{eq:V4}
\end{align}
This expression must be positive for all allowed values of $r$,
 $\zeta$ and $\xi$. Consider first $r$: from equation
\eqref{eq:definition_r} it is clear that $r$ can take any non-negative
value which means that, given the quadratic dependence of equation
\eqref{eq:V4} on $r$ that one must have
\begin{align}
0 & <\lambda_{H}\,,\label{eq:6}\\
0 & <\lambda_{\Delta}+\lambda'_{\Delta}\zeta\equiv F_{1}\left(\zeta\right)\,,\label{eq:7}\\
0 & <\lambda_{H\Delta}+\xi\lambda'_{H\Delta}+\sqrt{\lambda_{H}\left(\lambda_{\Delta}+\lambda'_{\Delta}\zeta\right)}\equiv F_{2}\left(\xi,\zeta\right)\,.\label{eq:8}
\end{align}
These conditions match those given in \cite{Arhrib:2011uy} with a
different notation. However, what follows differs with
\cite{Arhrib:2011uy} in a crucial way.

In order to obtain the necessary and sufficient conditions for the
quartic couplings $\lambda_{i}$ which yield a potential bounded from
below, one needs to get rid of $\zeta$ and $\xi$ from conditions
\eqref{eq:6}--\eqref{eq:8}. Note that these conditions must be
respected for all $\zeta$ and $\xi$, so one needs to find what are the
allowed values of $\left(\xi,\zeta\right)$ from the definition of
these two quantities. We do not show the details here, but the reader
can convince her/himself that $\xi$ can take any value between $0$ and
$1$ and $\zeta$ can be anywhere between $1/2$ and 1, as noted in
\cite{Arhrib:2011uy}.

\begin{figure}[tbph]
\begin{centering}
\includegraphics[scale=0.45]{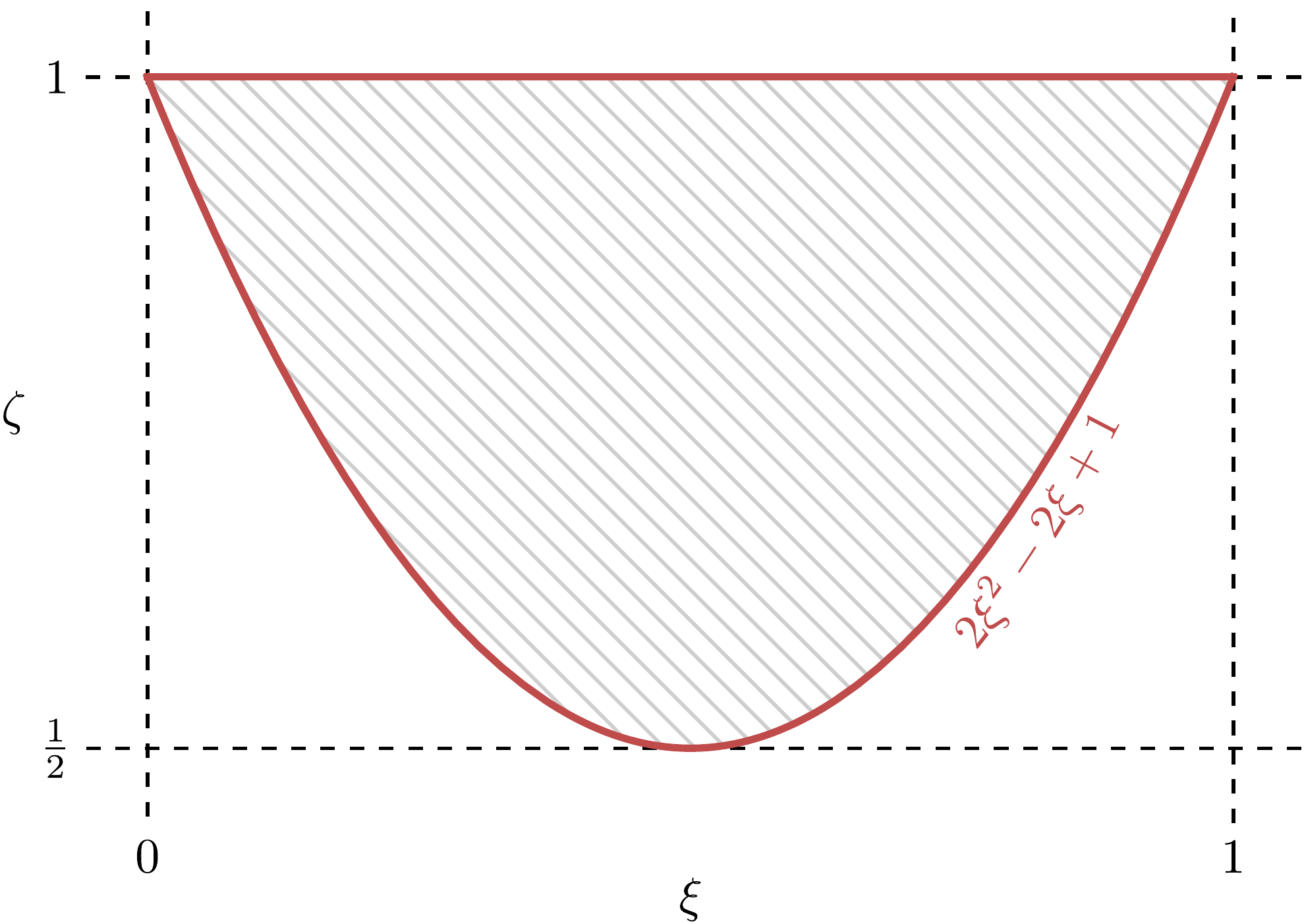}
\par\end{centering}
\protect\caption{\label{fig:1}The shaded region is the allowed one for
  the parameters $\left(\xi,\zeta\right)$.}
\end{figure}

However, the crucial point is that this does not mean that
$\left(\xi,\zeta\right)$ can be anywhere in the rectangle with
vertices in $\left(0,\frac{1}{2}\right)$ and
$\left(1,1\right)$. Indeed, from equations \eqref{eq:definition_rho}
and \eqref{eq:definition_phi} it can be shown that the possible values
of $\left(\xi,\zeta\right)$ correspond to
\begin{equation}
2\xi^{2}-2\xi+1\leq\zeta\leq1\,,
\end{equation}
which defines the shaded region depicted in figure \ref{fig:1}. 
Since the function $F_{1}\left(\zeta\right)$ defined in \eqref{eq:7}
is monotonic, the condition `$0<F_{1}\left(\zeta\right)$ for all
$\zeta$' is equivalent to `$0<F_{1}\left(\frac{1}{2}\right)$ and
$0<F_{1}\left(1\right)$' which translates into the requirement
\begin{equation}
0<\lambda_{\Delta}+\frac{1}{2}\lambda'_{\Delta}\textrm{ and }0<\lambda_{\Delta}+\lambda'_{\Delta}\,.
\end{equation}

As for the condition in \eqref{eq:8}, note that 
`$0<F_{2}\left(\xi,\zeta\right)$ for all $\xi$ and $\zeta$' is
trivially the same as
$0<\min F_{2}\left(\xi,\zeta\right)$, so one is left with
the job of finding the minimum of $F_{2}$. Furthermore, since this
function is monotonic in both $\xi$ and $\zeta$, we know that its
minimum occurs at the border of the shaded region in figure
\ref{fig:1}; to be more specific, this argument shows that the minimum
of the function must occur somewhere along the line defined by
$\zeta=2\xi^{2}-2\xi+1$, with $0\leq\xi\leq1$. Then we may take
\begin{align}
\widehat{F}\left(\xi\right) & \equiv F_{2}\left(\xi,2\xi^{2}-2\xi+1\right)
\end{align}
noticing that the sign of $\widehat{F}''\left(\xi\right)$ is constant
--- it is the same as the one of $\lambda'_{\Delta}$. Therefore, one
can always find a value $\xi_{0}$ where
$\widehat{F}'\left(\xi_{0}\right)=0$.  Such a $\xi_{0}$ will be a
minimum if $\widehat{F}''\left(\xi_{0}\right)>0$ and, furthermore, one
must also make sure that $0\leq\xi_{0}\leq1$ (or equivalently that
$\widehat{F}'\left(0\right)<0$ and $\widehat{F}'\left(1\right)>0$
since $\widehat{F}'$ is a monotonous function). This will be true if
and only if
$\lambda'_{\Delta}\sqrt{\lambda_{H}}>\left|\lambda'_{H\Delta}\right|\sqrt{\lambda_{\Delta}+\lambda'_{\Delta}}$,
in which case
\begin{align}
\widehat{F}\left(\xi_{0}\right) & =\lambda_{H\Delta}+\frac{1}{2}\lambda'_{H\Delta}+\frac{1}{2}\sqrt{\left(2\lambda_{H}\lambda'_{\Delta}-{\lambda'_{H\Delta}}^{2}\right)\left(2\frac{\lambda_{\Delta}}{\lambda'_{\Delta}}+1\right)}\,.
\end{align}
The remaining possibility is that the minimum of $\widehat{F}$ in the
interval $\xi\in\left[0,1\right]$ is at $\xi=0$ or $1$, from which we
get the constraints that both
$\widehat{F}\left(0\right)=\lambda_{H\Delta}+\sqrt{\lambda_{H}\left(\lambda_{\Delta}+\lambda'_{\Delta}\right)}$
and
$\widehat{F}\left(1\right)=\lambda_{H\Delta}+\lambda'_{H\Delta}+\sqrt{\lambda_{H}\left(\lambda_{\Delta}+\lambda'_{\Delta}\right)}$
should be positive quantities.

In summary, the potential will be bounded from below if and only if

\begin{gather}
\lambda_{H},\,\lambda_{\Delta}+\lambda'_{\Delta},\,\lambda_{\Delta}+\frac{1}{2}\lambda'_{\Delta},\,\lambda_{H\Delta}+\sqrt{\lambda_{H}\left(\lambda_{\Delta}+\lambda'_{\Delta}\right)},\,\lambda_{H\Delta}+\lambda'_{H\Delta}+\sqrt{\lambda_{H}\left(\lambda_{\Delta}+\lambda'_{\Delta}\right)}>0\nonumber \\
\textrm{and}\nonumber \\
\left[\lambda'_{\Delta}\sqrt{\lambda_{H}}\leq\left|\lambda'_{H\Delta}\right|\sqrt{\lambda_{\Delta}+\lambda'_{\Delta}}\textrm{ or }2\lambda_{H\Delta}+\lambda'_{H\Delta}+\sqrt{\left(2\lambda_{H}\lambda'_{\Delta}-{\lambda'_{H\Delta}}^{2}\right)\left(2\frac{\lambda_{\Delta}}{\lambda'_{\Delta}}+1\right)}>0\right]\,.\label{eq:bfb_condition}
\end{gather}

\begin{center}
\begin{figure}[!h]
\begin{centering}
\includegraphics[scale=0.35]{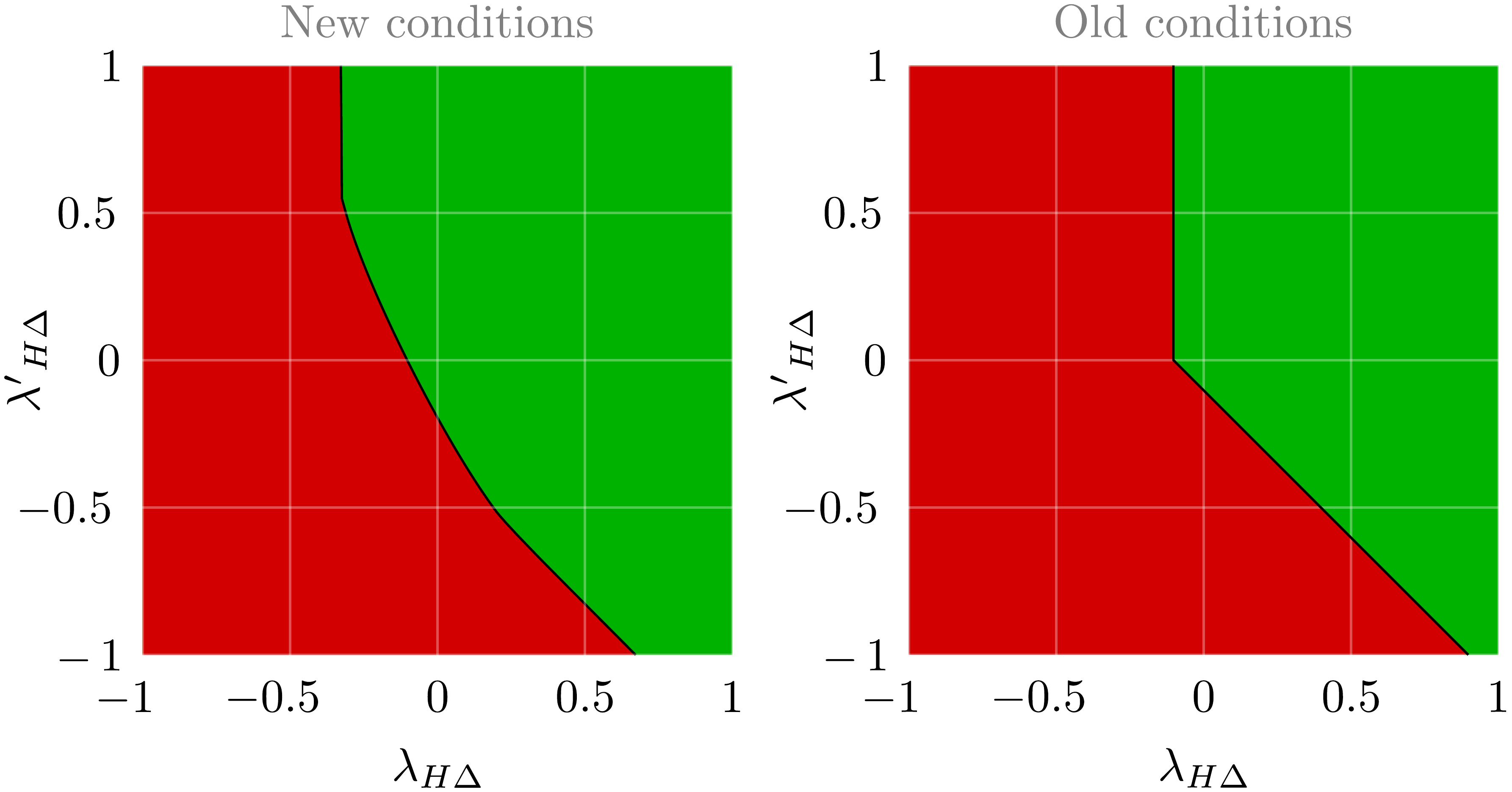}
\par\end{centering}

\protect\caption{\label{fig:stable_regions_no_running}Regions of
  stability (green) and instability (red) of the potential for
  $\lambda_{\Delta}=-\frac{1}{3}$, $\lambda'_{\Delta}=\frac{3}{4}$ and
  $\lambda_{H}\approx\frac{1}{4}$.  The two plots make it possible to
  compare the correct stability conditions as given in equation
  \eqref{eq:bfb_condition_literature} (left) with the ones in use in the
  literature (right).}
\end{figure}
\par\end{center}

The condition in \eqref{eq:bfb_condition} should be compared with the
one used up to now in the literature, where the last line of
\eqref{eq:bfb_condition} is replaced by
$F_{2}\left(0,\frac{1}{2}\right),\,F_{2}\left(1,\frac{1}{2}\right)>0$,
which translates into
\begin{equation}
\lambda_{H\Delta}+\sqrt{\lambda_{H}\left(\lambda_{\Delta}+\frac{1}{2}\lambda'_{\Delta}\right)},\,\lambda_{H\Delta}+\lambda'_{H\Delta}+\sqrt{\lambda_{H}\left(\lambda_{\Delta}+\frac{1}{2}\lambda'_{\Delta}\right)}>0\,.\label{eq:bfb_condition_literature}
\end{equation}
From the discussion so far it should be clear that this condition is
too strict: potentials $V$ which obey it are necessarily bounded from
below, but not all potentials which are bounded from below do obey
it. Indeed, the constraint in \eqref{eq:bfb_condition_literature}
assumes that by varying the fields $H$ and $\Delta$ the point
$\left(\xi,\zeta\right)$ can be anywhere within the dashed rectangle
in figure \ref{fig:1}, when in reality only the shaded region is
allowed, with two thirds of the area of the rectangle. Restricting to
the 5-dimensional box region where $\left|\lambda_{i}\right|\leq1$, a
numerical scan indicates that close to 5\% of the valid points are
excluded by the constraint in \eqref{eq:bfb_condition_literature},
although in certain special scenarios, as in figure
\ref{fig:stable_regions_no_running}, this percentage can be
significantly larger.

\section{\noindent\label{sec:Regions-of-stability}Regions of stability and perturbativity}

Now that we have the correct stability conditions we consider the
renormalization group evolution of the triplet seesaw model.
Ignoring all Yukawa couplings except the one of the top, using
\cite{Machacek:1983tz,Machacek:1983fi,Machacek:1984zw} one finds the
renormalization group equations of the model to be the following (see
also \cite{Chao:2006ye,Schmidt:2007nq}):\footnote{Using the dictionary
  in appendix \hyperref[sec:Appendix-A:-Conversion]{A}, it can be
  checked that these expressions match those in (3.2) of
  \cite{Chun:2012jw}, the only difference being that in
  $\left(4\pi\right)^{2}\frac{d\lambda_{4}}{dt}$, instead of a term
  $+\frac{9}{5}{g'}^{2}$, we get $+3{g'}^{2}$.}

\begin{align}
\left(4\pi\right)^{2}\frac{dg_{i}}{dt} & =b_{i}g_{i}^{3}\textrm{ with }b_{i}=\left(\frac{47}{10},-\frac{5}{2},-7\right)\,,\\
\left(4\pi\right)^{2}\frac{d\lambda_{H}}{dt} & =\frac{27}{100}g_{1}^{4}+\frac{9}{10}g_{1}^{2}g_{2}^{2}+\frac{9}{4}g_{2}^{4}-\left(\frac{9}{5}g_{1}^{2}+9g_{2}^{2}\right)\lambda_{H}+12\lambda_{H}^{2}+6\lambda_{H\Delta}^{2}+6\lambda_{H\Delta}\lambda'_{H\Delta}+\frac{5}{2}{\lambda'_{H\Delta}}^{2}\nonumber \\
 & +12\lambda_{H}y_{t}^{2}-12y_{t}^{4}\,,\\
\left(4\pi\right)^{2}\frac{d\lambda_{H\Delta}}{dt} & =\frac{27}{25}g_{1}^{4}-\frac{18}{5}g_{1}^{2}g_{2}^{2}+6g_{2}^{4}-\left(\frac{9}{2}g_{1}^{2}+\frac{33}{2}g_{2}^{2}\right)\lambda_{H\Delta}+6\lambda_{H}\lambda_{H\Delta}+2\lambda_{H}\lambda'_{H\Delta}+4\lambda_{H\Delta}^{2}\nonumber \\
 & +8\lambda_{\Delta}\lambda_{H\Delta}+6\lambda'_{\Delta}\lambda_{H\Delta}+{\lambda'_{H\Delta}}^{2}+3\lambda_{\Delta}\lambda'_{H\Delta}+\lambda'_{\Delta}\lambda'_{H\Delta}+6\lambda_{H\Delta}y_{t}^{2}\,,\\
\left(4\pi\right)^{2}\frac{d\lambda'_{H\Delta}}{dt} & =\frac{36}{5}g_{1}^{2}g_{2}^{2}-\left(\frac{9}{2}g_{1}^{2}+\frac{33}{2}g_{2}^{2}\right)\lambda'_{H\Delta}+2\lambda_{H}\lambda'_{H\Delta}+8\lambda_{H\Delta}\lambda'_{H\Delta}+4{\lambda'_{H\Delta}}^{2}+2\lambda_{\Delta}\lambda'_{H\Delta}\nonumber \\
 & +4\lambda'_{\Delta}\lambda'_{H\Delta}+6\lambda'_{H\Delta}y_{t}^{2}\,,\\
\left(4\pi\right)^{2}\frac{d\lambda_{\Delta}}{dt} & =\frac{108}{25}g_{1}^{4}-\frac{72}{5}g_{1}^{2}g_{2}^{2}+30g_{2}^{4}-\left(\frac{36}{5}g_{1}^{2}+24g_{2}^{2}\right)\lambda_{\Delta}+4\lambda_{H\Delta}^{2}+4\lambda_{H\Delta}\lambda'_{H\Delta}\nonumber \\
 & +14\lambda_{\Delta}^{2}+12\lambda_{\Delta}\lambda'_{\Delta}+3{\lambda'_{\Delta}}^{2}\,,\\
\left(4\pi\right)^{2}\frac{d\lambda'_{\Delta}}{dt} & =\frac{144}{5}g_{1}^{2}g_{2}^{2}-12g_{2}^{4}+2{\lambda'_{H\Delta}}^{2}-\left(\frac{36}{5}g_{1}^{2}+24g_{2}^{2}\right)\lambda'_{\Delta}+12\lambda_{\Delta}\lambda'_{\Delta}+9{\lambda'_{\Delta}}^{2}\,.
\end{align}

Using these equations and requiring stability of the scalar potential
in the energy range going from the top mass all the way to the Planck
mass one obtains the regions of quartic couplings indicated in green
in figure \ref{fig:stable_regions_upto_mPlanck}. The right panel
corresponds to the use of  the stability conditions used in the
literature, while the left panel refers to our new and less
restrictive stability conditions. On the other hand the instability
regions are indicated in red. Finally those cases which correspond to
a stable vacuum but involve non-perturbative dynamics because
$\left|\lambda_{i}\right|>\sqrt{4\pi}$ for some quartic coupling
$\lambda_{i}$ are indicated in orange.
Notice also that the stable region becomes bigger if one imposes
stability only up to some intermediate scale, chosen to be $10^{12}$
GeV, as indicated by the light green region in figure
\ref{fig:stable_regions_upto_mPlanck}.

\begin{center}
\begin{figure}
\begin{centering}
\includegraphics[scale=0.35]{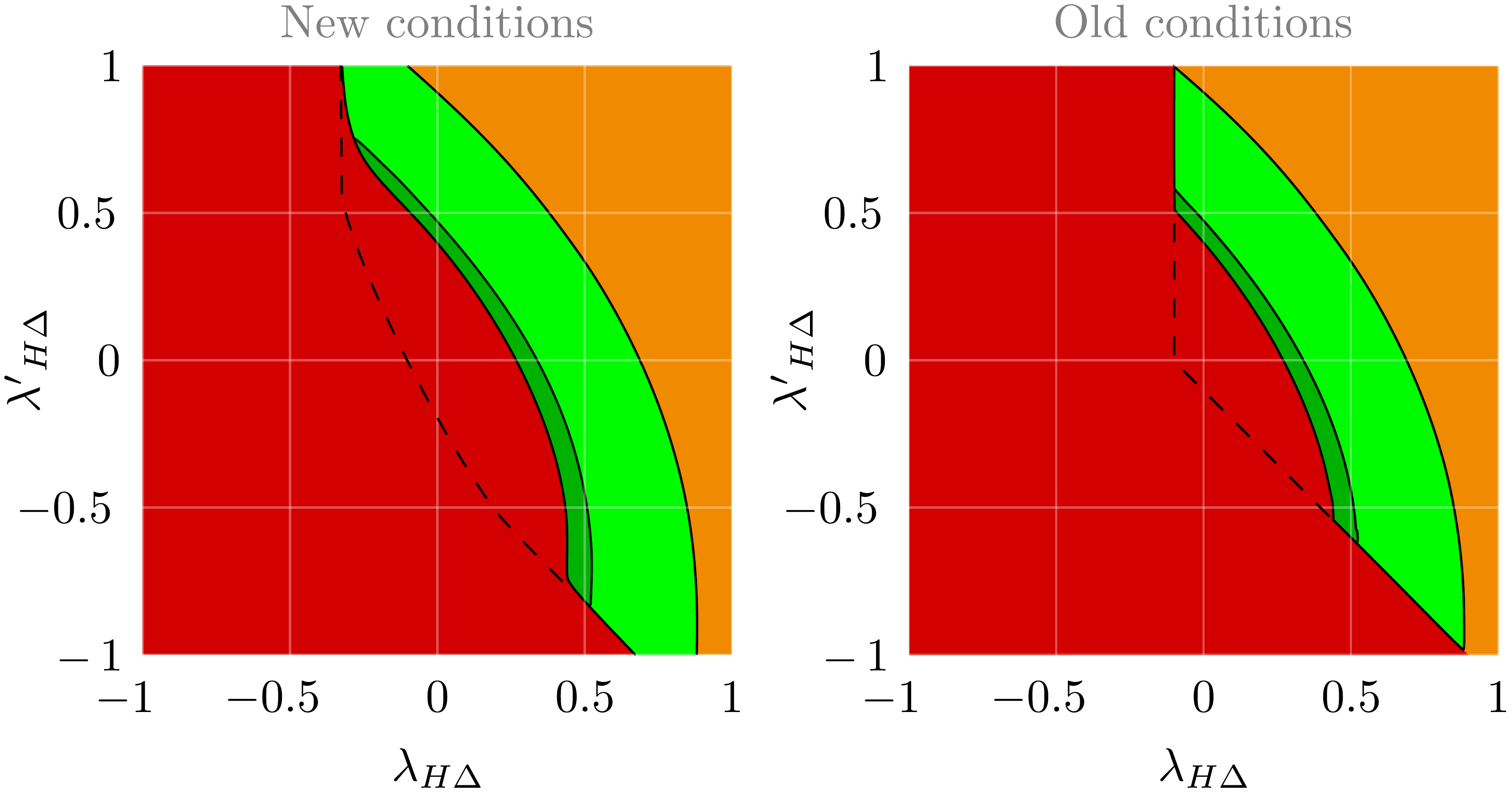}
\par\end{centering}

\protect\caption{\label{fig:stable_regions_upto_mPlanck}Regions of
  stability (dark green) and instability (red) considering the energy
  range going from the top mass all the way to the Planck mass. Those
  cases which (appear to) lead to a stable vacuum but involve
  non-perturbative dynamics because
  $\left|\lambda_{i}\right|>\sqrt{4\pi}$ for some quartic coupling
  $\lambda_{i}$ are shown in orange. If one requires stability only up
  to $10^{12}$ GeV the stable region becomes bigger, as indicated by
  the light green region. The dashed lines indicate the border between the stable and unstable regions at low energies (see figure \ref{fig:stable_regions_no_running}).}
\end{figure}
\par\end{center}

\section{\label{sec:Final-remarks} Phenomenological profile of the
  triplet seesaw Higgs sector}

%%\begin{shaded}{\color{tempText} 

Since its original proposal there have been many phenomenological
studies of the scalar sector of the triplet model, as it constitutes
an essential ingredient of the type-II seesaw mechanism. 
For the benefit of the reader we present in figure \ref{fig:artist} of
Appendix~\hyperref[sec:Appendix-B:-Artist]{B} a schematic view of the
scalar boson mass spectrum in the model given in
table~\ref{tab:Mass-eigenstates}.
One sees that, in addition to the SM Higgs boson found, one has heavy
neutral ($H^0$,$A^0$), singly ($H^+$) and doubly charged ($H^{++}$)
scalar bosons, whose mass is controlled by $\chi$ and with a small
splitting which should not be bigger than indicated on figure
\ref{fig:deltaM} if the model is to remain perturbative all the way up
to the Planck scale.

The doubly--charged state comes just from the triplet, while all other
heavy states come mainly from the triplet, but with a small admixture
with the \sm Higgs boson, controlled by the ratio of VEVs  $\alpha\equiv v_{\Delta}/v_{H}$.
Note that the state $A^{0}$ is identified with the would-be triplet
Nambu-Goldstone boson associated to spontaneous \lnv which
becomes massless as $v_{\Delta} \to 0$.
All of these scalar states have a nearly common mass, with a small
splitting, both indicated in figure \ref{fig:artist}. This follows
from the consistency requirements such as perturbativity studied in
the previous section and displayed in figure \ref{fig:deltaM}.  Hence,
altogether, once the lightest Higgs boson discovered at the LHC is
accommodated, one can describe fairly well the scalar sector with just three parameters ($\alpha$, $\lambda'_{H\Delta}$ and $\chi$).
This is in sharp contrast with other extended electroweak breaking
potentials, such as those of supersymmetric models.

For example the singly and doubly--charged members of the triplet have
been searched for at accelerators such as LEP as well as hadron
colliders~ \cite{Accomando:2006ga,Akeroyd:2005gt,ATLAS:2012hi,Chatrchyan:2012ya}.
If sufficiently light, say below 400 GeV or so, the $H^{++}$ will be
copiously produced at the LHC, which could enable interesting
measurements of its branching ratios of the various leptonic decay
channels~\cite{Akeroyd:2007zv}, as well as the leading $WW$ decay
branch~\cite{Kanemura:2014ipa,Kanemura:2014goa}. The former are
determined by the triplet Yukawa couplings. These determine also the
pattern of \lfv decays. Given the small neutrino masses indicated by
experiment~\cite{Forero:2014bxa,Nunokawa:2007qh,Barabash:2011fn,Deppisch:2012nb}
and our assumption that the scalars are in the TeV region, these
Yukawa couplings are expected to be too small to cause detectable
signals.

The near degeneracy of the heavy scalars implies that, once the
constraints on the charged Higgs bosons are imposed, by choosing a
suitably large $\chi$, the neutral ones, including the would--be
Majoron, will also have escaped detection at LEP.
Moreover, the charged components in the Higgs triplet model provide a
potential enhancement of the $H\to \gamma\gamma$ decay branching~\cite{
Arhrib:2011vc,Akeroyd:2012ms,Chun:2012jw}
ratio, which can be probed at the LHC.
Last but not least, the triplet introduces changes to the  $S,T,U$
oblique parameters.\footnote{In practice these are
  expected to be small, just like the changes in the $\rho$ parameter
  discussed previously.}

All of the above phenomena should be studied within parameter regions
where the electroweak symmetry breaking is consistent and, as we saw
in figure \ref{fig:stable_regions_upto_mPlanck}, consistency implies
strong restrictions on quartic parameter values.
Although the relevant restrictions apply mainly to the quartic scalar
interactions, and in principle do not translate directly into
stringent constraints upon the Higgs boson masses, one has an
important restriction on the splitting between the masses of the heavy
states, such as the singly and doubly charged scalar bosons,
illustrated by the funnel region depicted in figure \ref{fig:deltaM}.
Performing a dedicated phenomenological study of the scalar sector
lies outside the scope of this paper but we hope to have given a
helpful guideline.

One last word regarding the naturalness of the scalar potential in the
presence of the cubic mass parameter. This follows from the principle
that its removal would lead to a theory of enhanced symmetry, in which
neutrinos would be massless and lepton number would be conserved. In
any case, a dynamical completion of this theory in which the cubic
term is replaced by a quartic one is possible and has in fact been
suggested long ago~\cite{Schechter:1981cv}. This would imply the
presence of a mainly singlet Nambu-Goldstone boson with implications
for Higgs decays such as invisibly decaying Higgs
bosons~\cite{joshipura:1992hp,Diaz:1997xv,Diaz:1998zg} whose detailed
analysis is more general than the one recently given in reference
\cite{Bonilla2015} and lies outside the scope of the present paper.

\begin{center}
\begin{figure}
\begin{centering}
\includegraphics[scale=0.6]{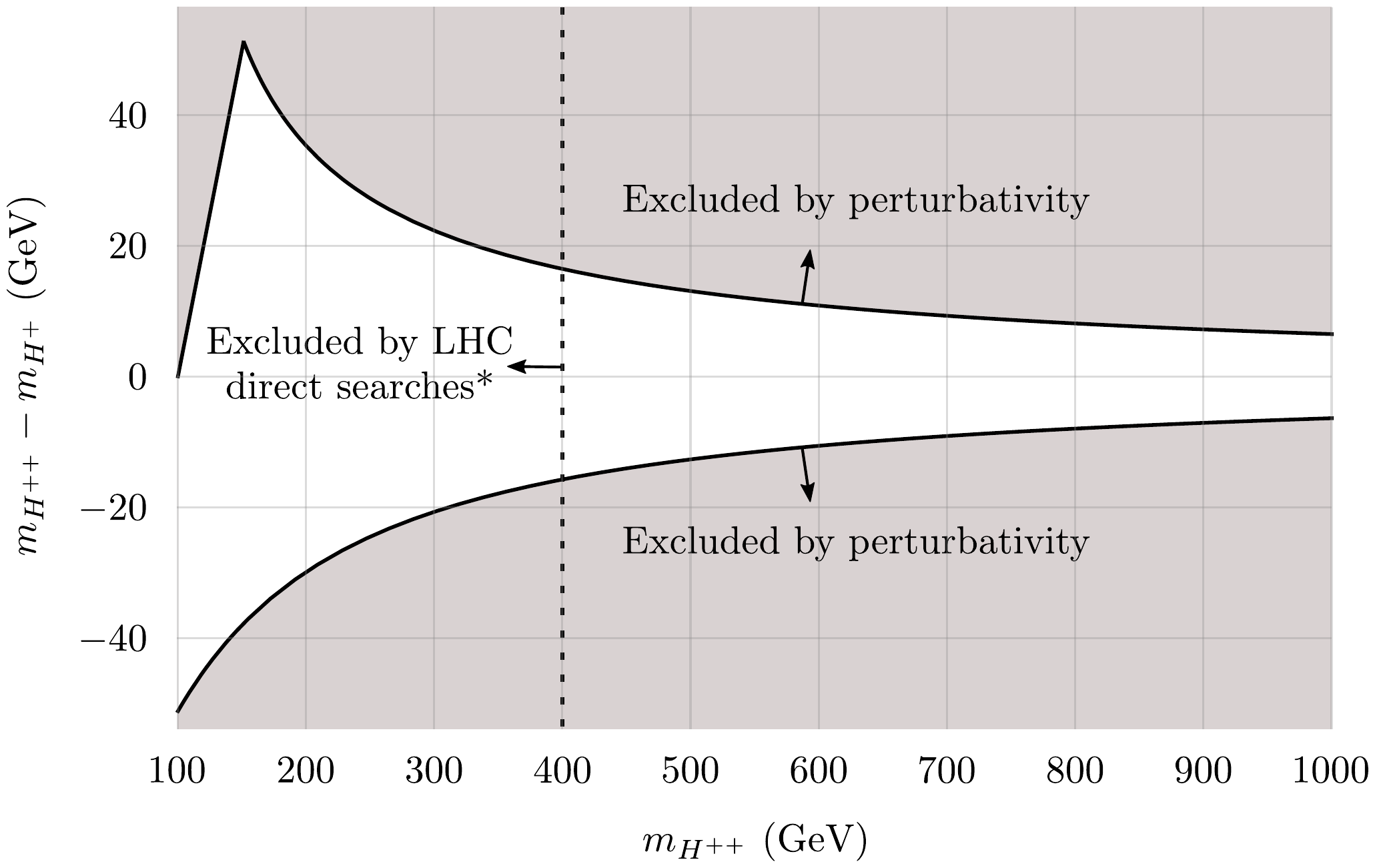}
\par\end{centering}

\protect\caption{\label{fig:deltaM}The coupling $\lambda'_{H\Delta}$
  must be roughly between -0.85 and 0.85 if all quartic couplings are
  to remain small up to $m_{Planck}$
  ($\left|\lambda_{i}\right|<\sqrt{4\pi}$). This perturbativity
  requirement strongly constrains the mass splitting of the triplet
  components, particularly if one considers the LHC lower bound
  $m_{H^{++}}\sim400$ GeV from direct searches of $H^{++}$ decaying in
  to leptons \cite{ATLAS:2012hi,Chatrchyan:2012ya} ({*}assuming 100\%
  branching fractions).  This plot also assumes that $m_{H^{+}}>100$
  GeV.}
\end{figure}

\par\end{center}

\section{\label{sec:Final-remarks}Final remarks}

In this paper, we have considered the consistency of the type-II
seesaw model symmetry breaking. We included under consistency both the
requirements of boundedness from below as well as perturbativity up to
some scale.  We found that the bounded-from-below conditions for the
scalar potential in use in the literature are not correct. For
definiteness and simplicity we focused on the case of explicit
violation of lepton number. We discussed some scenarios where the
correction we have found can be significant. Moreover we have sketched
the typical scalar boson profile expected by consistency of the
vacuum.
Before closing we note that, the restrictions discussed in this paper
do not depend on the hypercharge of the scalar triplet $\Delta$, hence the same set of
conditions also applies for any other model which extends the scalar sector of
the Standard Model with an $SU(2)_L$ triplet.

\subsection*{\vspace{-0.15cm}Acknowledgments}
This work supported by the Spanish grants FPA2014-58183-P, Multidark
CSD2009-00064 and SEV-2014-0398 (MINECO), and PROMETEOII/2014/084
(Generalitat Valenciana).

\newpage
\section*{\label{sec:Appendix-A:-Conversion} 
  Appendix A: Conversion between different notations}

Given that different notations are used in the literature to write
down the different terms in the scalar potential of the model, we
provide here table \ref{tab:Translation} to facilitate comparisons.
\begin{table}[!h]
\begin{centering}
\begin{tabular}{ccccccccc}
\toprule 
Source & $\mu_{H}^{2}$ & $\mu_{\Delta}^{2}$ & $m_{H\Delta}$ & $\lambda_{H}$ & $\lambda_{H\Delta}$ & $\lambda'_{H\Delta}$ & $\lambda_{\Delta}$ & $\lambda'_{\Delta}$\tabularnewline
\midrule
\cite{Arhrib:2011uy} & $m_{H}^{2}$ & $M_{\Delta}^{2}$ & $\frac{1}{2}\mu$ & $\frac{1}{2}\lambda$ & $\lambda_{1}$ & $\lambda_{4}$ & $2\lambda_{2}$ & $2\lambda_{3}$\tabularnewline
\cite{Chun:2012jw}{*} & $-m^{2}$ & $M^{2}$ & $\sqrt{2}\mu$ & $2\lambda_{1}$ & $\lambda_{4}-\lambda_{5}$ & $2\lambda_{5}$ & $2\lambda_{2}+2\lambda_{3}$ & $-2\lambda_{3}$\tabularnewline
\cite{Chao:2006ye}{*} & $-m_{\phi}^{2}$ & $M_{\xi}^{2}$ & $-\left(\lambda_{H}M_{\xi}\right)^{*}$ & $\frac{1}{2}\lambda$ & $\lambda_{\phi}-\frac{1}{2}\lambda_{T}$ & $\lambda_{T}$ & $4\lambda_{C}+\frac{1}{2}\lambda_{\xi}$ & $-4\lambda_{C}$\tabularnewline
\cite{Dev:2013ff}{*} & $-m_{\Phi}^{2}$ & $M_{\Delta}^{2}$ & $\sqrt{2}\Lambda_{6}$ & $\lambda$ & $\lambda_{4}+\lambda_{5}$ & $-2\lambda_{5}$ & $\lambda_{1}+\lambda_{2}$ & $-\lambda_{2}$\tabularnewline
\bottomrule
\end{tabular}
\par\end{centering}

\protect\caption{
  \label{tab:Translation}Translation between the notation used in this
  paper and the one used by other authors. Note that in the cases marked
  with an asterisk it is also necessary to flip the sign of the doubly
  charged component of the triplet.}
\end{table}

\section*{\label{sec:Appendix-B:-Artist} Appendix B:
  Representative triplet seesaw scalar mass spectrum}

In order to grasp in a visual manner the scalar spectrum of the model
(see table~\ref{tab:Mass-eigenstates}) as well as the effect on the
degeneracy of the three new scalars of having $\lambda'_{H\Delta}$
constrained to be roughly between -0.85 and 0.85, we present here
figure \ref{fig:artist}.
\begin{center}
\begin{figure}[!h]
\begin{centering}
\includegraphics[scale=0.8]{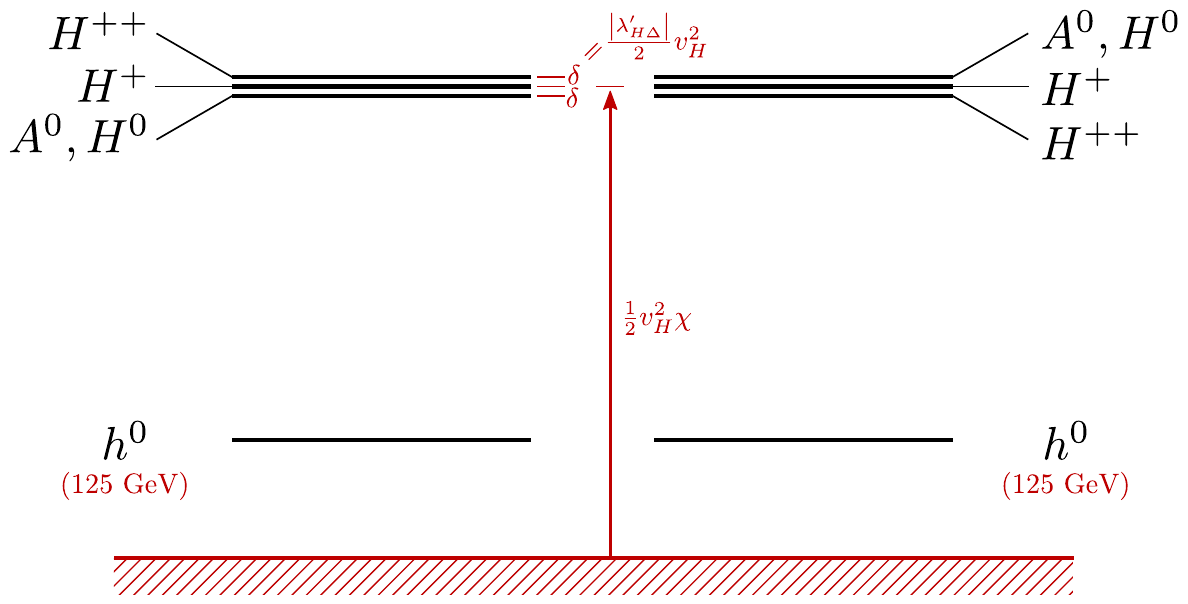}
\par\end{centering}

\protect\caption{\label{fig:artist} Schematic view of the scalar boson
  mass spectrum in the triplet seesaw model. The heavy scalars are
  nearly degenerate. The ordering of the heavy scalar masses depends on the sign of $\lambda'_{H\Delta}$ as shown in table~\ref{tab:Mass-eigenstates}. Recall that $\chi$ refers to the ratio $m_{H\Delta}/v_{\Delta}$.}
\end{figure}

\par\end{center}


\begin{thebibliography}{10}
\providecommand{\url}[1]{\texttt{#1}}
\providecommand{\urlprefix}{URL }
\providecommand{\eprint}[2][]{\url{#2}}

\bibitem{weinberg:1980bf}
S.~Weinberg, \emph{{Varieties of baryon and lepton nonconservation}},
  \MYhref[journalLinks]{http://dx.doi.org/10.1103/PhysRevD.22.1694}{Phys. Rev.
  }\MYhref[journalLinks]{http://dx.doi.org/10.1103/PhysRevD.22.1694}{\textbf{D22}
  (1980) 1694}.

\bibitem{Schechter:1980gr}
J.~Schechter and J.~Valle, \emph{{Neutrino masses in $SU(2) \times U(1)$
  theories}},
  \MYhref[journalLinks]{http://dx.doi.org/10.1103/PhysRevD.22.2227}{Phys.Rev.
  }\MYhref[journalLinks]{http://dx.doi.org/10.1103/PhysRevD.22.2227}{\textbf{D22}
  (1980) 2227}.

\bibitem{Schechter:1981cv}
J.~Schechter and J.~W.~F. Valle, \emph{{Neutrino decay and spontaneous
  violation of lepton number}},
  \MYhref[journalLinks]{http://dx.doi.org/10.1103/PhysRevD.25.774}{Phys. Rev.
  }\MYhref[journalLinks]{http://dx.doi.org/10.1103/PhysRevD.25.774}{\textbf{D25}
  (1982) 774}.

\bibitem{Sher:1988mj}
M.~Sher, \emph{{Electroweak Higgs potentials and vacuum stability}},
  \MYhref[journalLinks]{http://dx.doi.org/10.1016/0370-1573(89)90061-6}{Phys.
  Rept.
  }\MYhref[journalLinks]{http://dx.doi.org/10.1016/0370-1573(89)90061-6}{\textbf{179}
  (1989) 273--418}.

\bibitem{Alekhin:2012py}
S.~Alekhin, A.~Djouadi and S.~Moch, \emph{{The top quark and Higgs boson masses
  and the stability of the electroweak vacuum}},
  \MYhref[journalLinks]{http://dx.doi.org/10.1016/j.physletb.2012.08.024}{Phys.
  Lett.
  }\MYhref[journalLinks]{http://dx.doi.org/10.1016/j.physletb.2012.08.024}{\textbf{B716}
  (2012) 214--219},
  \MYhref[eprintLinks]{http://arxiv.org/abs/1207.0980}{{\ttfamily
  arXiv:1207.0980 [hep-ph]}}.

\bibitem{Gogoladze2008}
I.~Gogoladze, N.~Okada and Q.~Shafi, \emph{{Higgs boson mass bounds in a type
  II seesaw model with triplet scalars}},
  \MYhref[journalLinks]{http://dx.doi.org/10.1103/PhysRevD.78.085005}{Phys.
  Rev.
  }\MYhref[journalLinks]{http://dx.doi.org/10.1103/PhysRevD.78.085005}{\textbf{D78}
  (2008) 085005},
  \MYhref[eprintLinks]{http://arxiv.org/abs/0802.3257}{{\ttfamily
  arXiv:0802.3257 [hep-ph]}}.

\bibitem{Chao2012}
W.~Chao, M.~Gonderinger and M.~J. Ramsey-Musolf, \emph{{Higgs vacuum stability,
  neutrino mass, and dark matter}},
  \MYhref[journalLinks]{http://dx.doi.org/10.1103/PhysRevD.86.113017}{Phys.
  Rev.
  }\MYhref[journalLinks]{http://dx.doi.org/10.1103/PhysRevD.86.113017}{\textbf{D86}
  (2012) 113017},
  \MYhref[eprintLinks]{http://arxiv.org/abs/1210.0491}{{\ttfamily
  arXiv:1210.0491 [hep-ph]}}.

\bibitem{Chun:2012jw}
E.~J. Chun, H.~M. Lee and P.~Sharma, \emph{{Vacuum stability, perturbativity,
  EWPD and Higgs-to-diphoton rate in type II seesaw models}},
  \MYhref[journalLinks]{http://dx.doi.org/10.1007/JHEP11(2012)106}{JHEP
  }\MYhref[journalLinks]{http://dx.doi.org/10.1007/JHEP11(2012)106}{\textbf{11}
  (2012) 106}, \MYhref[eprintLinks]{http://arxiv.org/abs/1209.1303}{{\ttfamily
  arXiv:1209.1303 [hep-ph]}}.

\bibitem{Dev:2013ff}
P.~Bhupal~Dev, D.~K. Ghosh, N.~Okada and I.~Saha, \emph{{125 GeV Higgs boson
  and the type-II seesaw model}},
  \MYhref[journalLinks]{http://dx.doi.org/10.1007/JHEP03(2013)150,
  10.1007/JHEP05(2013)049}{JHEP
  }\MYhref[journalLinks]{http://dx.doi.org/10.1007/JHEP03(2013)150,
  10.1007/JHEP05(2013)049}{\textbf{1303} (2013) 150},
  \MYhref[eprintLinks]{http://arxiv.org/abs/1301.3453}{{\ttfamily
  arXiv:1301.3453 [hep-ph]}}.

\bibitem{Lebedev2012}
O.~Lebedev, \emph{{On stability of the electroweak vacuum and the Higgs
  portal}},
  \MYhref[journalLinks]{http://dx.doi.org/10.1140/epjc/s10052-012-2058-2}{Eur.
  Phys. J.
  }\MYhref[journalLinks]{http://dx.doi.org/10.1140/epjc/s10052-012-2058-2}{\textbf{C72}
  (2012) 2058}, \MYhref[eprintLinks]{http://arxiv.org/abs/1203.0156}{{\ttfamily
  arXiv:1203.0156 [hep-ph]}}.

\bibitem{Elias-Miro2012}
J.~Elias-Miro et~al., \emph{{Stabilization of the electroweak vacuum by a
  scalar threshold effect}},
  \MYhref[journalLinks]{http://dx.doi.org/10.1007/JHEP06(2012)031}{JHEP
  }\MYhref[journalLinks]{http://dx.doi.org/10.1007/JHEP06(2012)031}{\textbf{06}
  (2012) 031}, \MYhref[eprintLinks]{http://arxiv.org/abs/1203.0237}{{\ttfamily
  arXiv:1203.0237 [hep-ph]}}.

\bibitem{Falkowski2015}
A.~Falkowski, C.~Gross and O.~Lebedev, \emph{{A second Higgs from the Higgs
  portal}},
  \MYhref[journalLinks]{http://dx.doi.org/10.1007/JHEP05(2015)057}{JHEP
  }\MYhref[journalLinks]{http://dx.doi.org/10.1007/JHEP05(2015)057}{\textbf{05}
  (2015) 057}, \MYhref[eprintLinks]{http://arxiv.org/abs/1502.01361}{{\ttfamily
  arXiv:1502.01361 [hep-ph]}}.

\bibitem{Costa:2014qga}
R.~Costa, A.~P. Morais, M.~O.~P. Sampaio and R.~Santos, \emph{{Two-loop
  stability of a complex singlet extended Standard Model}},
  \MYhref[journalLinks]{http://dx.doi.org/10.1103/PhysRevD.92.025024}{Phys.
  Rev.
  }\MYhref[journalLinks]{http://dx.doi.org/10.1103/PhysRevD.92.025024}{\textbf{D92}
  (2015) 2 025024},
  \MYhref[eprintLinks]{http://arxiv.org/abs/1411.4048}{{\ttfamily
  arXiv:1411.4048 [hep-ph]}}.

\bibitem{Bonilla2015}
C.~Bonilla, R.~M. Fonseca and J.~W.~F. Valle, \emph{{Vacuum stability with
  spontaneous violation of lepton number}}  (2015),
  \MYhref[eprintLinks]{http://arxiv.org/abs/1506.04031}{{\ttfamily
  arXiv:1506.04031 [hep-ph]}}.

\bibitem{Arhrib:2011uy}
A.~Arhrib et~al., \emph{{The Higgs potential in the type II seesaw model}},
  \MYhref[journalLinks]{http://dx.doi.org/10.1103/PhysRevD.84.095005}{Phys.Rev.
  }\MYhref[journalLinks]{http://dx.doi.org/10.1103/PhysRevD.84.095005}{\textbf{D84}
  (2011) 095005},
  \MYhref[eprintLinks]{http://arxiv.org/abs/1105.1925}{{\ttfamily
  arXiv:1105.1925 [hep-ph]}}.

\bibitem{Agashe:2014kda}
K.~Olive et~al. (Particle Data Group), \emph{{Review of particle physics}},
  \MYhref[journalLinks]{http://dx.doi.org/10.1088/1674-1137/38/9/090001}{Chin.Phys.
  }\MYhref[journalLinks]{http://dx.doi.org/10.1088/1674-1137/38/9/090001}{\textbf{C38}
  (2014) 090001}.

\bibitem{Machacek:1983tz}
M.~E. Machacek and M.~T. Vaughn, \emph{{Two loop renormalization group
  equations in a general quantum field theory. 1. Wave function
  renormalization}},
  \MYhref[journalLinks]{http://dx.doi.org/10.1016/0550-3213(83)90610-7}{Nucl.
  Phys.
  }\MYhref[journalLinks]{http://dx.doi.org/10.1016/0550-3213(83)90610-7}{\textbf{B222}
  (1983) 83}.

\bibitem{Machacek:1983fi}
M.~E. Machacek and M.~T. Vaughn, \emph{{Two loop renormalization group
  equations in a general quantum field theory. 2. Yukawa couplings}},
  \MYhref[journalLinks]{http://dx.doi.org/10.1016/0550-3213(84)90533-9}{Nucl.
  Phys.
  }\MYhref[journalLinks]{http://dx.doi.org/10.1016/0550-3213(84)90533-9}{\textbf{B236}
  (1984) 221}.

\bibitem{Machacek:1984zw}
M.~E. Machacek and M.~T. Vaughn, \emph{{Two loop renormalization group
  equations in a general quantum field theory. 3. Scalar quartic couplings}},
  \MYhref[journalLinks]{http://dx.doi.org/10.1016/0550-3213(85)90040-9}{Nucl.
  Phys.
  }\MYhref[journalLinks]{http://dx.doi.org/10.1016/0550-3213(85)90040-9}{\textbf{B249}
  (1985) 70}.

\bibitem{Chao:2006ye}
W.~Chao and H.~Zhang, \emph{{One-loop renormalization group equations of the
  neutrino mass matrix in the triplet seesaw model}},
  \MYhref[journalLinks]{http://dx.doi.org/10.1103/PhysRevD.75.033003}{Phys.
  Rev.
  }\MYhref[journalLinks]{http://dx.doi.org/10.1103/PhysRevD.75.033003}{\textbf{D75}
  (2007) 033003},
  \MYhref[eprintLinks]{http://arxiv.org/abs/hep-ph/0611323}{{\ttfamily
  arXiv:hep-ph/0611323 [hep-ph]}}.

\bibitem{Schmidt:2007nq}
M.~A. Schmidt, \emph{{Renormalization group evolution in the type I+II seesaw
  model}}, \MYhref[journalLinks]{http://dx.doi.org/10.1103/PhysRevD.85.099903,
  10.1103/PhysRevD.76.073010}{Phys. Rev.
  }\MYhref[journalLinks]{http://dx.doi.org/10.1103/PhysRevD.85.099903,
  10.1103/PhysRevD.76.073010}{\textbf{D76} (2007) 073010}, [Erratum: Phys.
  Rev.D85,099903(2012)],
  \MYhref[eprintLinks]{http://arxiv.org/abs/0705.3841}{{\ttfamily
  arXiv:0705.3841 [hep-ph]}}.

\bibitem{Accomando:2006ga}
\emph{{Workshop on CP studies and non-standard Higgs physics}} (2006),
  \MYhref[eprintLinks]{http://arxiv.org/abs/hep-ph/0608079}{{\ttfamily
  arXiv:hep-ph/0608079 [hep-ph]}}.

\bibitem{Akeroyd:2005gt}
A.~G. Akeroyd and M.~Aoki, \emph{{Single and pair production of doubly charged
  Higgs bosons at hadron colliders}},
  \MYhref[journalLinks]{http://dx.doi.org/10.1103/PhysRevD.72.035011}{Phys.
  Rev.
  }\MYhref[journalLinks]{http://dx.doi.org/10.1103/PhysRevD.72.035011}{\textbf{D72}
  (2005) 035011},
  \MYhref[eprintLinks]{http://arxiv.org/abs/hep-ph/0506176}{{\ttfamily
  arXiv:hep-ph/0506176}}.

\bibitem{ATLAS:2012hi}
G.~Aad et~al. (ATLAS), \emph{{Search for doubly-charged Higgs bosons in
  like-sign dilepton final states at $\sqrt{s}=7$ TeV with the ATLAS
  detector}},
  \MYhref[journalLinks]{http://dx.doi.org/10.1140/epjc/s10052-012-2244-2}{Eur.
  Phys. J.
  }\MYhref[journalLinks]{http://dx.doi.org/10.1140/epjc/s10052-012-2244-2}{\textbf{C72}
  (2012) 2244}, \MYhref[eprintLinks]{http://arxiv.org/abs/1210.5070}{{\ttfamily
  arXiv:1210.5070 [hep-ex]}}.

\bibitem{Chatrchyan:2012ya}
S.~Chatrchyan et~al. (CMS), \emph{{A search for a doubly-charged Higgs boson in
  $pp$ collisions at $\sqrt{s}=7$ TeV}},
  \MYhref[journalLinks]{http://dx.doi.org/10.1140/epjc/s10052-012-2189-5}{Eur.
  Phys. J.
  }\MYhref[journalLinks]{http://dx.doi.org/10.1140/epjc/s10052-012-2189-5}{\textbf{C72}
  (2012) 2189}, \MYhref[eprintLinks]{http://arxiv.org/abs/1207.2666}{{\ttfamily
  arXiv:1207.2666 [hep-ex]}}.

\bibitem{Akeroyd:2007zv}
A.~G. Akeroyd, M.~Aoki and H.~Sugiyama, \emph{{Probing Majorana phases and
  neutrino mass spectrum in the Higgs triplet model at the LHC}},
  \MYhref[journalLinks]{http://dx.doi.org/10.1103/PhysRevD.77.075010}{Phys.
  Rev.
  }\MYhref[journalLinks]{http://dx.doi.org/10.1103/PhysRevD.77.075010}{\textbf{D77}
  (2008) 075010},
  \MYhref[eprintLinks]{http://arxiv.org/abs/0712.4019}{{\ttfamily
  arXiv:0712.4019 [hep-ph]}}.

\bibitem{Kanemura:2014ipa}
S.~Kanemura, M.~Kikuchi, H.~Yokoya and K.~Yagyu, \emph{{LHC Run-I constraint on
  the mass of doubly charged Higgs bosons in the same-sign diboson decay
  scenario}}, \MYhref[journalLinks]{http://dx.doi.org/10.1093/ptep/ptv071}{PTEP
  }\MYhref[journalLinks]{http://dx.doi.org/10.1093/ptep/ptv071}{\textbf{2015}
  (2015) 051B02},
  \MYhref[eprintLinks]{http://arxiv.org/abs/1412.7603}{{\ttfamily
  arXiv:1412.7603 [hep-ph]}}.

\bibitem{Kanemura:2014goa}
S.~Kanemura, M.~Kikuchi, K.~Yagyu and H.~Yokoya, \emph{{Bounds on the mass of
  doubly-charged Higgs bosons in the same-sign diboson decay scenario}},
  \MYhref[journalLinks]{http://dx.doi.org/10.1103/PhysRevD.90.115018}{Phys.
  Rev.
  }\MYhref[journalLinks]{http://dx.doi.org/10.1103/PhysRevD.90.115018}{\textbf{D90}
  (2014) 11 115018},
  \MYhref[eprintLinks]{http://arxiv.org/abs/1407.6547}{{\ttfamily
  arXiv:1407.6547 [hep-ph]}}.

\bibitem{Forero:2014bxa}
D.~Forero, M.~Tortola and J.~Valle, \emph{{Neutrino oscillations refitted}},
  \MYhref[journalLinks]{http://dx.doi.org/10.1103/PhysRevD.90.093006}{Phys.Rev.
  }\MYhref[journalLinks]{http://dx.doi.org/10.1103/PhysRevD.90.093006}{\textbf{D90}
  (2014) 9 093006},
  \MYhref[eprintLinks]{http://arxiv.org/abs/1405.7540}{{\ttfamily
  arXiv:1405.7540 [hep-ph]}}.

\bibitem{Nunokawa:2007qh}
H.~Nunokawa, S.~J. Parke and J.~W. Valle, \emph{{CP violation and neutrino
  oscillations}},
  \MYhref[journalLinks]{http://dx.doi.org/10.1016/j.ppnp.2007.10.001}{Prog.Part.Nucl.Phys.
  }\MYhref[journalLinks]{http://dx.doi.org/10.1016/j.ppnp.2007.10.001}{\textbf{60}
  (2008) 338--402},
  \MYhref[eprintLinks]{http://arxiv.org/abs/0710.0554}{{\ttfamily
  arXiv:0710.0554 [hep-ph]}}.

\bibitem{Barabash:2011fn}
A.~Barabash, \emph{{75 years of double beta decay: yesterday, today and
  tomorrow}}  (2011),
  \MYhref[eprintLinks]{http://arxiv.org/abs/1101.4502}{{\ttfamily
  arXiv:1101.4502 [nucl-ex]}}.

\bibitem{Deppisch:2012nb}
F.~F. Deppisch, M.~Hirsch and H.~Pas, \emph{{Neutrinoless double beta decay and
  physics beyond the Standard Model}},
  \MYhref[journalLinks]{http://dx.doi.org/10.1088/0954-3899/39/12/124007}{J.Phys.
  }\MYhref[journalLinks]{http://dx.doi.org/10.1088/0954-3899/39/12/124007}{\textbf{G39}
  (2012) 124007},
  \MYhref[eprintLinks]{http://arxiv.org/abs/1208.0727}{{\ttfamily
  arXiv:1208.0727 [hep-ph]}}.

\bibitem{Arhrib:2011vc}
A.~Arhrib et~al., \emph{{Higgs boson decay into 2 photons in the type~II Seesaw
  Model}},
  \MYhref[journalLinks]{http://dx.doi.org/10.1007/JHEP04(2012)136}{JHEP
  }\MYhref[journalLinks]{http://dx.doi.org/10.1007/JHEP04(2012)136}{\textbf{04}
  (2012) 136}, \MYhref[eprintLinks]{http://arxiv.org/abs/1112.5453}{{\ttfamily
  arXiv:1112.5453 [hep-ph]}}.

\bibitem{Akeroyd:2012ms}
A.~G. Akeroyd and S.~Moretti, \emph{{Enhancement of H to gamma gamma from
  doubly charged scalars in the Higgs triplet model}},
  \MYhref[journalLinks]{http://dx.doi.org/10.1103/PhysRevD.86.035015}{Phys.
  Rev.
  }\MYhref[journalLinks]{http://dx.doi.org/10.1103/PhysRevD.86.035015}{\textbf{D86}
  (2012) 035015},
  \MYhref[eprintLinks]{http://arxiv.org/abs/1206.0535}{{\ttfamily
  arXiv:1206.0535 [hep-ph]}}.

\bibitem{joshipura:1992hp}
A.~S. Joshipura and J.~W.~F. Valle, \emph{{Invisible Higgs decays and neutrino
  physics}}, Nucl. Phys. \textbf{B397} (1993) 105--122.

\bibitem{Diaz:1997xv}
M.~A. Diaz, M.~A. Garcia-Jareno, D.~A. Restrepo and J.~W.~F. Valle,
  \emph{{Neutrino mass and missing momentum Higgs boson signals}},
  \MYhref[journalLinks]{http://dx.doi.org/10.1103/PhysRevD.58.057702}{Phys.
  Rev.
  }\MYhref[journalLinks]{http://dx.doi.org/10.1103/PhysRevD.58.057702}{\textbf{D58}
  (1998) 057702},
  \MYhref[eprintLinks]{http://arxiv.org/abs/hep-ph/9712487}{{\ttfamily
  arXiv:hep-ph/9712487 [hep-ph]}}.

\bibitem{Diaz:1998zg}
M.~A. Diaz, M.~A. Garcia-Jareno, D.~A. Restrepo and J.~W.~F. Valle,
  \emph{{Seesaw Majoron model of neutrino mass and novel signals in Higgs boson
  production at LEP}},
  \MYhref[journalLinks]{http://dx.doi.org/10.1016/S0550-3213(98)00434-9}{Nucl.
  Phys.
  }\MYhref[journalLinks]{http://dx.doi.org/10.1016/S0550-3213(98)00434-9}{\textbf{B527}
  (1998) 44--60},
  \MYhref[eprintLinks]{http://arxiv.org/abs/hep-ph/9803362}{{\ttfamily
  arXiv:hep-ph/9803362 [hep-ph]}}.

\end{thebibliography}
\end{document}